\documentclass[authoryear,preprint,review,12pt]{elsarticle}

\usepackage{lineno,hyperref}
\modulolinenumbers[5]
\usepackage{tabularx}

\newcommand\clearrow{\global\let\rowmac\relax}
\clearrow

\usepackage[T1]{fontenc}
\usepackage[utf8]{inputenc}
\usepackage{amsmath}
\usepackage{natbib}
\usepackage{graphicx}
\usepackage{gensymb}
\usepackage{url}
\usepackage{mathtools}
\usepackage[varg]{txfonts}
\usepackage{textcomp}
\usepackage[super]{nth}
\usepackage{caption}
\usepackage{subcaption}
\usepackage{lineno}
\usepackage{wasysym}
\usepackage{esvect}
\usepackage{ulem}

\usepackage{siunitx}
\usepackage{longtable}
\usepackage{lscape}
\usepackage{threeparttable}

\biboptions{authoryear}

\begin{document}

\begin{frontmatter}

\title{Correcting radar meteor fluxes for observing biases}

\author[uwopa,wiese]{Margaret Campbell-Brown}
\ead{margaret.campbell@uwo.ca}

\address[uwopa]{Department of Physics and Astronomy, University of Western Ontario, London, Ontario, N6A 3K7, Canada}
\address[wiese]{Western Institute for Earth and Space Exploration, University of Western Ontario, London, Ontario, N6A 5B7, Canada}

\begin{abstract}

We report on an eight year survey of simultaneous optical and radar meteor detections with the goal of isolating the fraction of meteors missed by specular radars.  A total of 10,503 Electron Multiplied Charge Couple Device (EMCCD) meteors with peak brightness above +7 were simultaneously detected by the Canadian Meteor Orbit Radar (CMOR) and used to estimate the fraction of radar echoes missed as a function of speed and height. During the time period that our cameras were recording, we found some 34,119 and 18,008 meteor echoes in total occurred within the field of view of the EMCCD cameras at 29 and 38 MHz respectively. This demonstrated that a significant fraction of the specular radar echoes remain below the detection threshold of the EMCCD cameras. We used these data to derive corrections for radar-specific observing biases. The optical height distributions for six velocity bins, spanning 10 - 72 km/s were used to determine the fraction of potential radar echoes which are observed for meteors with higher begin heights ($k_c$>96) and for those with lower begin heights. When compared to the correction factors currently used in calculating shower fluxes with CMOR, mid-velocity showers such as the Geminids and South Delta Aquariids are found to have been overcorrected, while high velocity showers like the Perseids and Leonids are significantly undercorrected. 

\end{abstract}

\end{frontmatter}

%\modulolinenumbers[1]
%\linenumbers

\section{Introduction}

Patrol meteor radars can provide fluxes for meteor showers in any weather conditions and during the day \citep[e.g.][]{Bruzzone2015,Campbell2021,Dewsnap2021}. Specular radars, however, are strongly affected by the height ceiling effect \citep{Greenhow1960}, in which echoes high in the atmosphere suffer from destructive interference due to their large widths and are not detected. This effectively limits the number of meteors which can be seen, with the effect being largest at short wavelengths. There are a number of other radar-specific observing biases related to the initial radius, including the finite velocity effect and the Pulse Repetition Factor (PRF) effect \citep{Ceplecha1998}. 

The initial radius of a meteor trail is effectively a function of the mean free path in the atmosphere; the trail will expand very rapidly for the first few collisions of meteor atoms with atmospheric molecules, because of the extreme relative speeds of the two populations \citep{BaggaleyWebb1977}. After of order 10 collisions \citep{Jones1995}, the ionized trail will thermalize, and then diffuse more slowly into the atmosphere.

Past studies of initial radius have used observations at multiple radar wavelengths to derive a correction factor. \citet{Greenhow1960} used simultaneous radar observations at 8.3 and 17 m (36 and 17.6 MHz, respectively), along with a height distribution measured at 4.3 m (70 MHz) to determine the ``true" height distribution of meteors and infer a correction for initial radius. They estimated that nearly all meteors were detected on the 17~m wavelength system at low speeds, falling to less than 50\% at high speeds; at 8.3~m the fraction varied from 40\% to less than 10\%. They used the echo diffusion time to calculate heights for the meteors, and Fresnel oscillations to calculate the speeds, which may bias the measurements. 

A number of other studies \citep{Kashcheyev1963,Baggaley1980,Elford1988} used two frequencies to characterize initial radius, but used amplitude ratios rather than height distributions. In principle, the amplitude ratio between echoes at a particular height should be equal to the ratio of the number of echoes seen at each frequency at that height. These studies also used decay times to calculate heights, and either Fresnel oscillations or the rise time of the echoes to calculate the speeds. 

\citet{Campbell2001} showed that amplitude ratios are strongly affected by meteoroid fragmentation. The amplitude ratio between echoes simultaneously observed at 29 and 38 MHz showed scatter at all heights of a factor of several, which could be explained by a spread of meteoroid fragments perpendicular to the direction of travel; interference among these fragments would be different on the different frequencies and was sufficient to explain the scatter. \citet{Jones2005} returned to assuming the shape of the trails was Gaussian in order to derive a correction for radar fluxes from the amplitude ratios, which is still in use for CMOR data.

The initial radius itself (as opposed to the correction to meteor rates) was examined using three frequency CMOR data by \citet{Stober2021}. They used a full wave scattering model with three different electron density profiles to calculate the initial radius, diffusion coefficient and electron line density. They found the Gaussian model worked best, but fewer than 5\% of the echoes could be fit on all three frequencies, most likely because of fragmentation. 

The finite velocity effect \citep{Galligan2001} takes into account the diffusion of the trail in the time the meteoroid takes to cross the first Fresnel zone, where most of the radar power is scattered. This zone is typically about 1 km in length. The correction is expected to be important at slow meteor speeds, where the time to traverse the first Fresnel zone is significant, but also at large heights, where diffusion is rapid in the low density atmosphere. 

The Pulse Repetition Factor (PRF) effect \citep{Galligan2001} specifically addresses the effects of diffusion on the detectability of meteors. A radar requires a certain number of radio pulse returns from a single meteor to distinguish between meteors and noise, so if the trail decays too rapidly, a detectable echo will not be identified as a meteor. This effect obviously depends on the number of pulses required for a meteor detection and the PRF of the radar. 

Because of fragmentation, using multi-frequency radar observations to calculate initial radius and other observing biases is difficult. In this work, we investigate destructive interference observing biases using the Canadian Meteor Orbit Radar (CMOR) and optical meteor observations to correct for the effects of initial radius, finite velocity and the PRF effect.

\section{Observations}

\subsection{Radar Observations}

The Canadian Meteor Orbit Radar (CMOR) is a three-frequency radar (17.45, 29.85, 38.15 MHz) located near Tavistock, Ontario, Canada (43.18653$^\circ$N, 80.67092$^\circ$W). The three systems use all-sky Yagi antennas to transmit and receive, and each have five receive antennas which can be used to locate echoes in the sky by interferometry \citep{Jones2005cmor, Brown2008}. The 17 MHz system suffers from interference, particularly during the day when the ionosphere is low. The 29 MHz system has higher power than the other two systems (15 kW compared to 6 kW for 17 and 38, during this observing period), and also five additional receivers located 5 - 22 km from the main site, which allow orbits to be determined for some echoes. The orbital data are not used for calculating shower fluxes, since the time-of-flight method used to determine speeds requires more signal to noise than single station observations, and the requirement for multiple specular scattering across sites for trajectory estimation biases certain geometries for detection. The correction for meteors undetected because orbits could not be calculated is uncertain. 

The radars have been running regularly since 2002 \citep{Brown2008}, with some equipment upgrades in 2011 \citep{Ye2013}. The three frequencies each have two detection pipelines running on them: the original Skiymet detection algorithm, which prioritizes strong underdense echoes suitable for wind measurements (called the mpd pipeline after the naming of the files) \citep{Hocking2001}, and the Western-developed pipeline, which uses much less stringent detection criteria to capture both overdense and very weak underdense echoes, saved as Meteor EVent files (called the mev pipeline) \citep{Weryk_Brown_2012, Mazur2020}. The mev pipeline is used for calculating mass indices, since the observations span a larger range of masses \citep{Blaauw2011}, but there is too much noise in the selected echoes (hence interferometry is often incorrect) to reliably calculate fluxes from this dataset. The analysis in this paper will focus on the mpd pipeline, used for fluxes.

\subsection{Electron-Multiplied-Charge Coupled Device (EMCCD) cameras}

The Canadian Automated Meteor Observatory (CAMO) \citep{Weryk2013} is a system of optical sensors designed to study very faint meteors. Among the optical systems comprising CAMO are a set of four EMCCD sensors, located in two pairs, one at Elginfield (43.1928$^\circ$N, 81.3157$^\circ$W), and the other co-located with CMOR near Tavistock \citep{Gural2022}. The cameras are Nuvu HNu1024 cameras, with 1024$\times$1024 pixels, with a frame rate of 16.7 fps; after March 2018 the cameras ran with 2$\times$2 binning, and the frame rate was increased to 32 fps. The cameras have a quantum efficiency of more than 90\%, and very low noise characteristics. The limiting stellar magnitude is normally +10, making these the most sensitive cameras in CAMO. The equivalent limiting peak meteor magnitude is approximately +7. All four cameras used 50~mm f/1.2 Nikkor lenses and had a roughly 15$\times$15$^\circ$ field of view. The two pairs of cameras were positioned so that the observed volumes were adjacent in the sky; one pair were positioned lower and one higher to reduce height bias. The cameras together have significant collecting area in the common volume from 70 to 150 km. 

The cameras ran in the above configuration from 2017 to 2024, and collected approximately 134,000 two-station meteors.

\section{Data reduction}

%Section on height distributions of three radars, and heights derived from decay times
\subsection{Radar height distributions}

\begin{figure}
  \includegraphics[width=\linewidth]{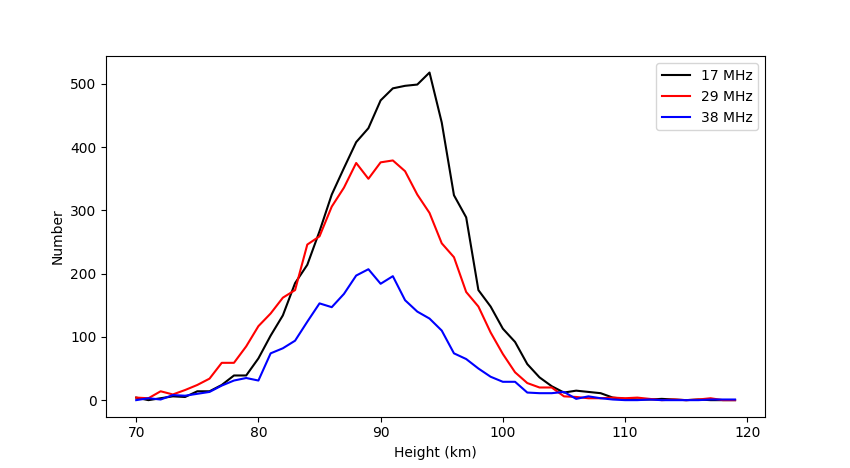}
  \caption[Height distributions of echoes on 17, 29 and 38 MHz CMOR systems for 2021 Jan 16] {Height distributions of echoes on 17, 29 and 38 MHz CMOR systems for 2021 Jan 16. There is no noticeable trend of higher echoes at lower frequencies. Note that the three frequencies do not have equal transmit power; the 29 MHz system is higher by a factor of more than 2.}
  \label{fig:htdists}
\end{figure}

As a first step, we attempted to recreate the \citet{Greenhow1960} height plots with our three frequencies. The plots were made for days when 17 MHz was relatively free of noise. Fig.~\ref{fig:htdists} shows the height distributions for a typical day (2021 Jan 16). The three curves overlap almost completely: there is a slight shift in the peak height, but there is no significant progression of the highest height observed, and certainly the distributions cannot be used to bootstrap the true height distribution. Every day of data examined had these characteristics. 

\begin{figure}
  \includegraphics[width=\linewidth]{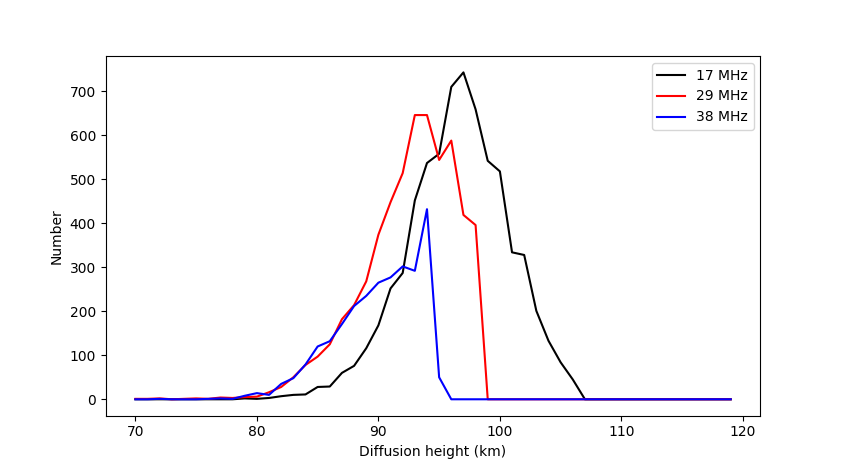}
  \caption[Decay time height distributions of echoes on 17, 29 and 38 MHz CMOR systems for 2021 Jan 16] {Decay time height distributions of echoes on 17, 29 and 38 MHz CMOR systems for 2021 Jan 16. When the heights are calculated using diffusion instead of interferometry, the lower frequencies show higher heights.}
  \label{fig:htdists_tau}
\end{figure}

Since \cite{Greenhow1960} did not have an interferometer to locate echoes in the sky, their heights were calculated from the decay time of the echo and a model of diffusion as a function of height. We used the same model on the calculated decay times of CMOR echoes to calculate heights, and redid the height distributions for all three frequencies (Fig.~\ref{fig:htdists_tau}). Using the diffusion to calculate heights shifts the distributions of the lower frequencies to higher heights. These are not correct, as we can see from the fact that simultaneous meteors captured on the three radars have interferometric heights which agree closely among the three systems. This result is not surprising as it has been recognized for some time that meteor echo based decay-time estimates of the ambipolar diffusion coefficient show considerable scatter and systematic uncertainty \citep[e.g.][]{Younger2014}.  It appears that the height distributions in \citet{Greenhow1960} were similarly biased, leading to an erroneous conclusion. 

\subsection{EMCCD-detected meteors specular to CMOR}

Since the three radar frequencies cannot be used to bootstrap a true height distribution, another method is needed. Here we make use of the fact that the EMCCD cameras can observe meteors that produce ionization trails which may be detected by CMOR.

The first step in this analysis was to find meteors observed by the EMCCD cameras which have the correct geometry to be observed by CMOR, in other words, which are specular to the line of sight of the radar.  Of the 134,000 meteors observed by more than one camera, just under 112,000 had good trajectory solutions, with speeds between 10 and 75 km/s, begin heights higher than their end heights, and occurred within a height window between 60 and 180 km. Of these, based on the two station optical trajectory solutions, approximately 14,000 were specular to the radar (within 2 degrees) at some point on their observed trajectory. The 2 degree limit was overly generous: more than 80\% of the meteors selected had angles between 89.5 and 90.5 degrees, and only those meteors beginning just after the specular point or ending just before it had more than half a degree difference from the right angle. Because we will calculate the radar correction factors as a function of meteoroid begin heights, we removed meteors which began outside the field of view of all four cameras. This left 10,503 meteors.

The single-station radar data from the mpd pipeline on all three frequencies was then searched for echoes at the appropriate time, range, and direction, within 2 seconds, 6 km in range (3 km being one range bin for CMOR) and 5 degrees in the angular coordinates. The rate of accidental association (when optical meteors from one day were sought in radar data from another day) was approximately 4 in 10,000. On 17 MHz, 4128 of the 10,503 potential meteors were observed; there were 4203 on 29 MHz and 2996 on 38 MHz. There were 304 EMCCD meteors observed at times when the radars were not running, and these were removed from the dataset. Note that the mev pipeline numbers of simultaneously observed meteors were 6133, 6372 and 5021 on 17, 29 and 38 MHz, consistent with the known looser detection criteria used for mev events compared to the mpd pipeline \citep{Mazur2020}. These associations will be used in future analyses; here we focus on mpd-optical associations only.

\begin{figure}
  \includegraphics[width=\linewidth]{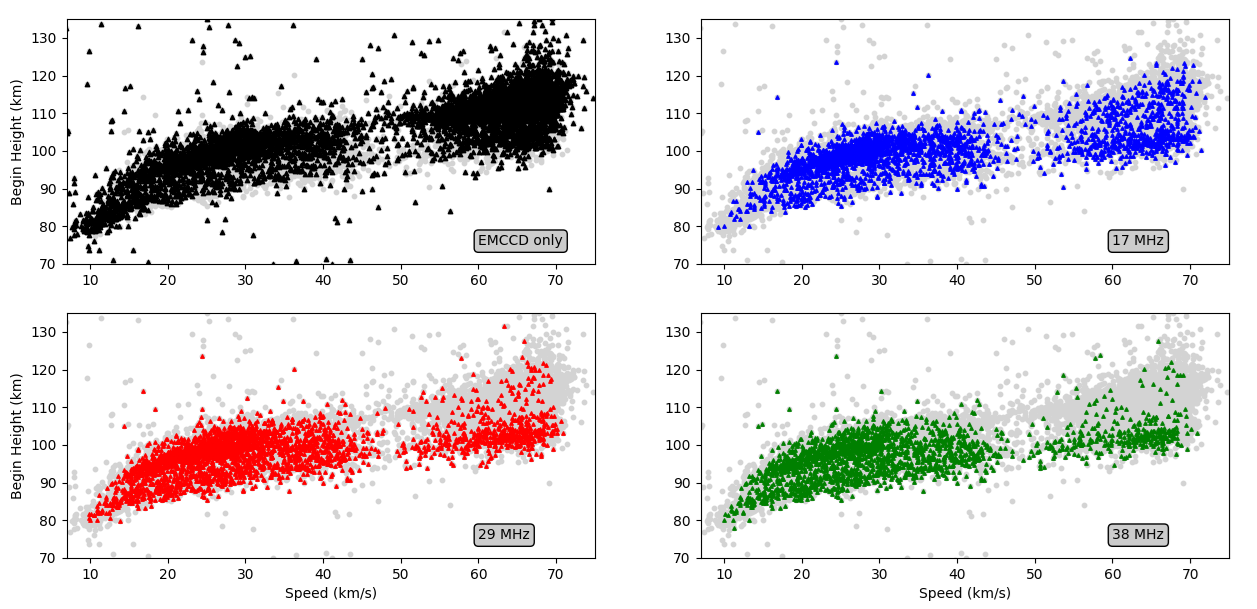}
  \caption[Begin heights and speeds of EMCCD meteors specular with CMOR]{The light gray in each plot are all the specular EMCCD meteors which could have been observed with CMOR. The upper left plot shows EMCCD meteors not observed by any frequency; the upper right overlays those meteors seen by the 17 MHz system; the lower left shows 29 MHz echoes and the lower right 38 MHz echoes.  Note the excess of meteors missed by the radars at the lowest speeds and heights, and also at the highest speeds and heights. The gap near 50~km s$^{-1}$ is the known minimum in the speed histogram between the slower, mostly prograde meteoroid population peaking around 30~km s$^{-1}$ and the faster, mainly retrograde population peaking around 65~km~s$^{-1}$} 
  \label{fig:HbV 3 freq}
\end{figure}

Figure \ref{fig:HbV 3 freq} shows the begin heights of these specular EMCCD meteors plotted against speed, overlaid with co-detections for each frequency of radar. While meteors are missed throughout the distribution, there is an excess of meteors not observed by the radars at the lowest speeds and low heights, and at the highest speeds and greatest heights. The two main populations of meteors are obvious in this plot: there is a high altitude population which corresponds to weak or volatile meteors which begin high in the atmosphere, and a low population which are stronger or more refractory and begin lower. These are the C and A populations, respectively, in the Ceplecha small meteoroid classification system \citep{Ceplecha1998}. 

\subsection{Height distributions}

\begin{figure}
  \includegraphics[width=\linewidth]{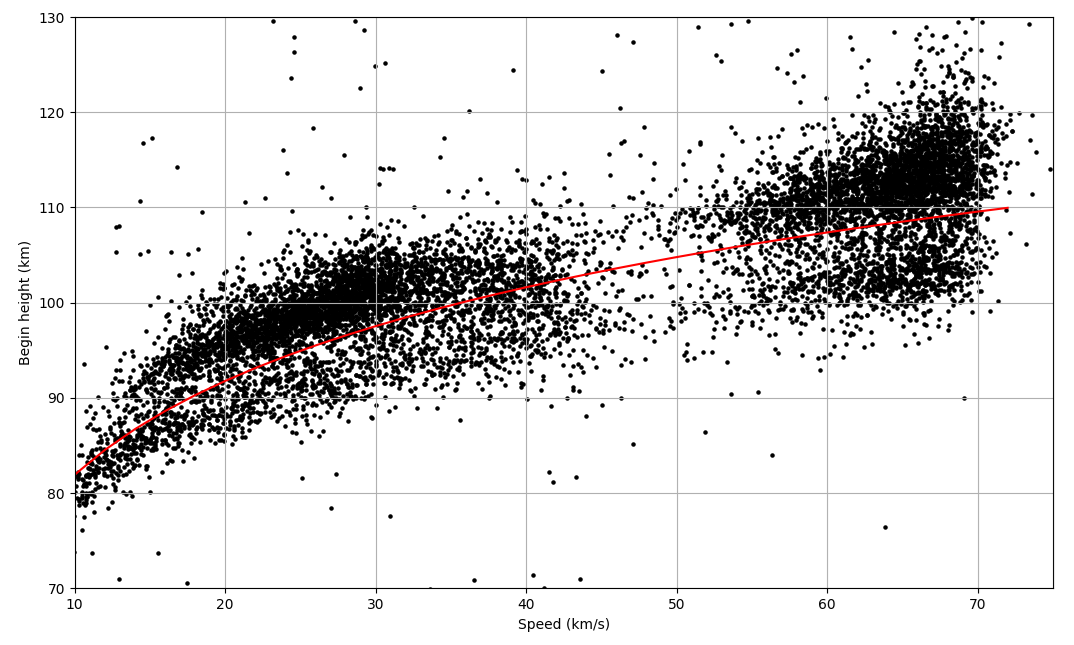}
  \caption[Separating meteors based on $k_c$ parameter]{The EMCCD meteors in our survey with specular geometry are plotted as begin height vs. speed. A $k_c$ parameter of 96 was selected to divide the high and low meteor populations, shown by the red line. }
  \label{fig:HbV_kc}
\end{figure}

The height of a meteor depends strongly on its speed. We divided the meteors into six velocity bins of 10 km/s width, starting at 10 km/s (meteors with speeds between 70 and 72 km/s were added to the last bin). However, there is still a wide range of heights in each speed bin, because of the diversity in physical and chemical properties of the meteoroids within each bin.  Fig.~\ref{fig:HbV_kc} shows the same begin height/speed plot as Fig.~\ref{fig:HbV 3 freq}, with all specular EMCCD meteors shown.  After some experimentation, we have chosen the $k_c$ parameter \citep{Jenniskens2016} to distinguish the high and low meteor populations, calculated using the EMCCD begin heights and speeds: $k_c=h_B+(2.86-2\log_{10}v)/0.0612$. A reference line at $k_c$=96 shows the division that was chosen. In general, our expectation is that the high population corresponds to weaker, less refractory meteoroids than the low population.

\begin{figure}
  \includegraphics[width=\linewidth]{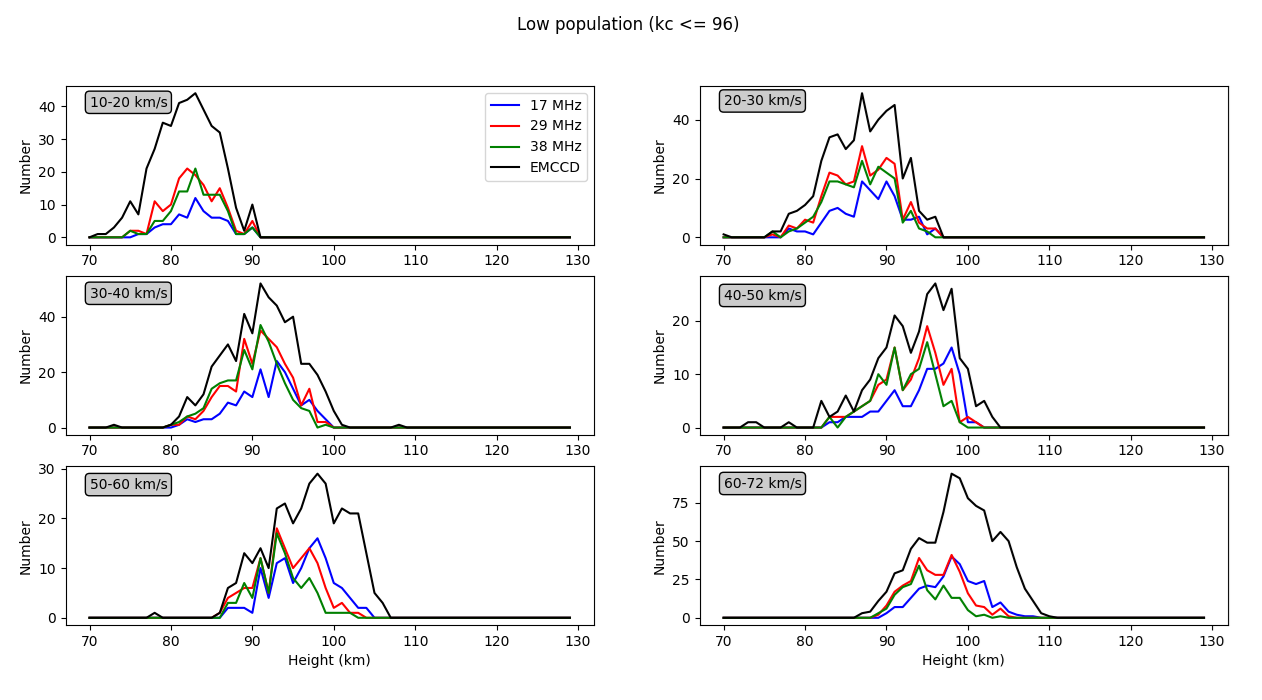}
  \caption[Low ($k_c<=96$) population height distributions]{Height distributions for the low ($k_c<=96$) population, binned by speed, for the specular EMCCD meteors, and from these specular EMCCD events those actually detected by each CMOR frequency. }
  \label{fig:hts_sim_low}
\end{figure}

\begin{figure}
  \includegraphics[width=\linewidth]{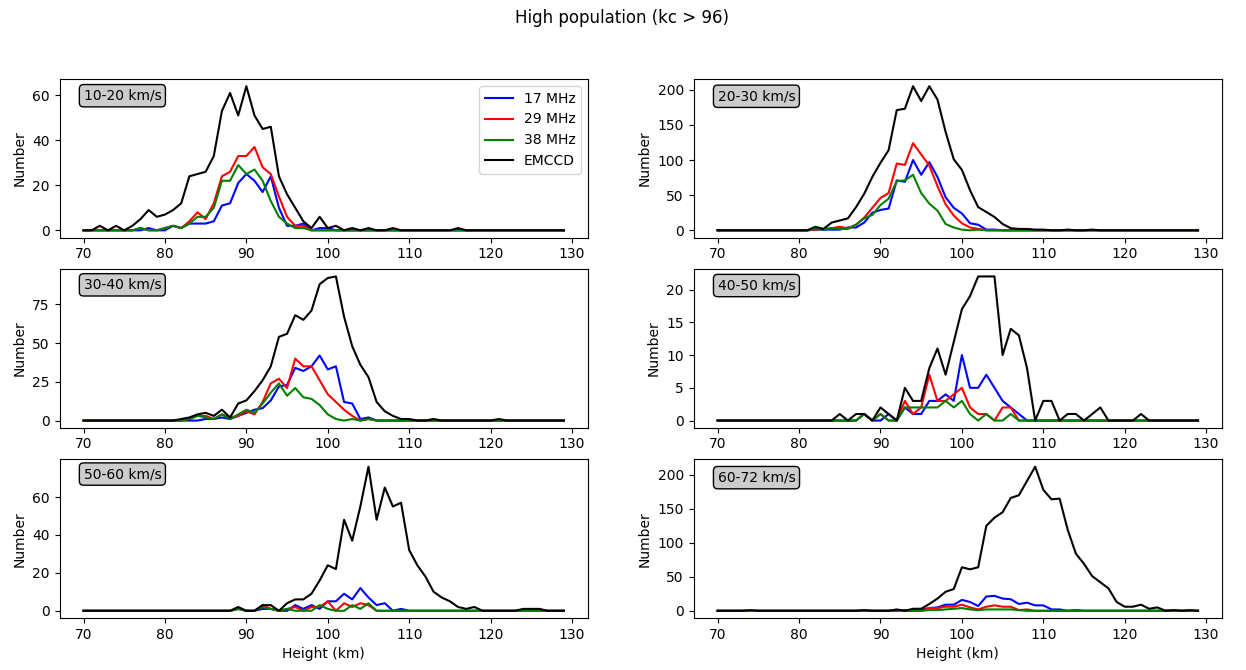}
  \caption[High ($k_c>96$) population height distributions]{Height distributions for the high ($k_c>96$) population, binned by speed, for the specular EMCCD meteors, and from these specular EMCCD events those actually detected by each CMOR frequency.}
  \label{fig:hts_sim_high}
\end{figure}

There are now twelve separate height distributions; six speed bins each for the high and low populations. Fig.~\ref{fig:hts_sim_low} shows height distributions for all the speed bins for the low ($k_c<=96$) population, and Fig.~\ref{fig:hts_sim_high} for the high population. The heights here are the heights of the specular point for each meteor, and the speeds are the speed at the time the meteor could have been observed by the radar, as opposed to the initial speed. Both height and speed are taken from the optical data.

The 17 MHz data here still suffers from missing data: removing times when the radars were off does not improve those results. The height distributions at 17 MHz are shown here, but they are not included in further analysis. Our inability to calculate correction factors is not a serious flaw, since the same noise issues make it impossible to calculate reliable shower fluxes at 17 MHz. 

\section{Results}

\subsection{New correction factors}

Using the height distributions above, we can correct the number of radar echoes at each frequency to the total number of meteors in that speed bin, as observed by the cameras. We are treating the set of specular EMCCD meteors as the complete observed population, and finding the correction factor for only these meteors. Note that the radars detect many more meteors in the EMCCD field of view than the cameras do: see section \ref{OverUnder}. We will assume that the fraction of meteors seen by the EMCCD cameras is the same as the fraction of fainter meteors not seen by the cameras.

To compare the fraction of echoes detected to the current flux correction factors, we combine the initial radius, finite velocity and PRF effect corrections. We use our current initial radius correction factor \citep{Jones2005}, an empirical correction which depends on the wavelength, the limiting magnitude, the speed, the zenith angle and the mass distribution index $s$. Each of these things affect the heights at which the meteors ablate, and therefore the amount of attenuation. We used the speed at the centre of each bin, a limiting magnitude for radar echoes of +8.1 (consistent with prior estimates for CMOR - see \citep{Brown2008, Weryk_Brown_2012} and a zenith angle of 60 degrees (the mean pointing direction of the centre of the EMCCD field of view). We have assumed an $s$ of 2.05 for the entire analysis, since the meteors in our dataset are mainly sporadics and this is the average sporadic mass index measured for CMOR \citep{Pokorny2016}.

The finite velocity correction uses a simple diffusion model to account for the increase in the ionized trail radius in the time it takes the meteor to cross the first Fresnel zone. It is calculated as follows ($k$ is the wavenumber, $2\pi/\lambda$; $D$ is the diffusion coefficient; $h$ the height and $R$ the range, both in meters; speed $v$ is in m/s):

$$D = 10^{0.00006h-4.74}$$
$$\Delta = \frac{2k^2D\sqrt{2R\lambda}}{v}$$
$$FV=\frac{1-\exp({-\Delta})}{\Delta}$$

Here, $h$ is calculated using the average height in each speed bin from the EMCCD meteor data.

The PRF effect is calculated from the radar parameters. CMOR requires a minimum of 5 pulses for detection, and has a PRF of 532 pulses per second. The correction factor is:

$$\omega = \frac{4 k^2D}{PRF}$$
$$PRF_{cor} = \frac{1-\exp({-\omega})}{\omega \exp({(n_{pulse}-1)\omega})}$$

\begin{figure}
  \includegraphics[width=\linewidth]{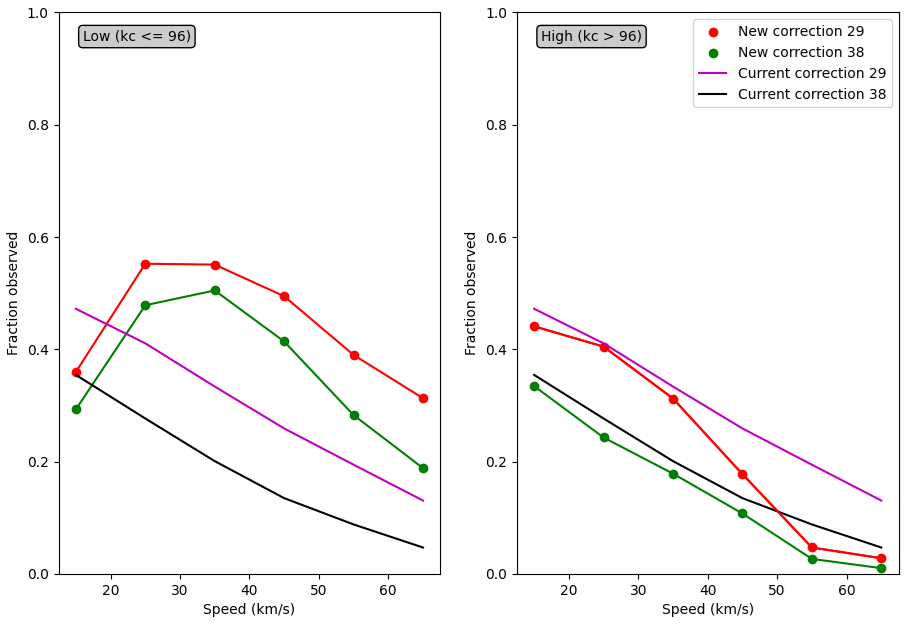}
  \caption[New correction factor]{Comparison of average correction currently used for initial radius, finite velocity and PRF effect with new correction factors for the high and low meteor populations. }
  \label{fig:ITR cor}
\end{figure}

Figure~\ref{fig:ITR cor} shows, for the low and high meteor populations, the new correction factors for 29 and 38 MHz compared to the average old correction factors. The lower population is overcorrected at almost all speeds except the slowest by the current corrections; the higher population correction changes significantly only for the fastest meteors, which are undercorrected at both frequencies, though especially so at 29 MHz. 

The current flux correction does not take into account slow meteors missed because of insufficient ionization when collision energies are low. The new analysis shows that this is very important for the low population at the lowest speed bins, where the trend of higher observed fraction increasing with decreasing speed reverses significantly. The effect is not strong in the high population. 

\subsection{Specific shower corrections}

It is interesting to see how much the correction factors change for specific major showers. To calculate the new corrections, we need to know whether to use the correction for the high or low population (or some combination). To do this, we look at showers observed with the EMCCD cameras on the same HbV (height by velocity) diagram as done previously for Figures~\ref{fig:HbV 3 freq}~and~\ref{fig:HbV_kc}. 

\begin{figure}
  \includegraphics[width=\linewidth]{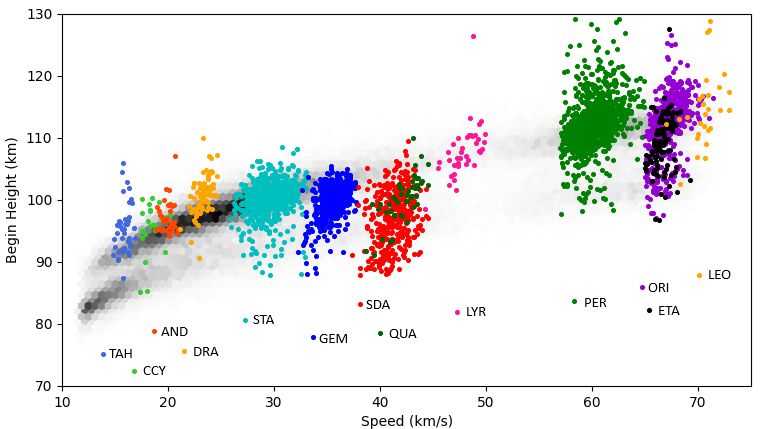}
  \caption[HbV diagram with showers]{Plot of begin height against speed of full EMCCD dataset (ie. not just those EMCCd meteors specular to CMOR) as a density plot, with selected shower meteors overplotted. }
  \label{fig:showers_HbV}
\end{figure}

Figure~\ref{fig:showers_HbV} shows the begin heights of selected showers in the full set of EMCCD meteors, represented as a density plot behind them. Here the shower events are identified using the criteria for shower membership and the same shower catalogue as used for the Global Meteor Network \citep{Vida2021}. The fast ($v>$45 km/s) and slower ($v<$25 km/s) showers belong to the high population, with a few members in the low population. The showers with moderate speeds, particularly the Geminids, South Delta Aquariids and Quadrantids, are located mainly between the high and low populations. The Draconids and Tau Herculids are especially high. 

Table~\ref{tab:ShowCorFacs} shows the selected showers, the number of shower members in the EMCCD dataset, the fraction of those meteors with a $k_c$<96, and the correction employed. We have chosen to use the high (or low) population when 75\% or more of the meteors belong to that population. Two showers, the Geminids and Eta Aquariids, are split nearly half and half between the populations, so we will use the average of the two correction factors.  

\begin{table}
		\centering
		\caption{Correction factors for showers}
		\label{tab:ShowCorFacs}
		\begin{tabular}{lccccc}
			\hline
			Shower  & Number & Percent in low             & Correction used & Change in & Change in\\
		       & & population & & correction 29 MHz & correction 38 MHz \\
            \hline
			TAH     & 39    & 2.6    & High  & 0.78 & 0.70   \\
			CCY     & 22    & 18.1    & High & 0.82 & 0.73 \\
			AND     & 33    & 0    & High & 0.89 & 0.88  \\
            DRA  & 75 & 2.7 & High & 0.90 & 0.88\\
            STA & 608 & 10.8 & High & 0.97 & 1.0\\
            GEM & 487 & 43.1 & (High + Low)/2 & 0.78 & 0.59\\
            SDA & 379 & 87.3 & Low & 0.60 & 0.39\\
            QUA & 47 & 78.7 & Low & 0.58 & 0.38\\
            LYR & 39 & 10.2 & High & 1.7 & 1.5\\
            PER & 966 & 6.5 & High & 5.1 & 4.6\\
            ORI & 610 & 15.1 & High & 5.7 & 6.2\\
            ETA & 106 & 44.3 & (High + Low)/s & 0.93 & 0.61\\
            LEO & 28 & 14.3 & High & 7.1 & 203\\
			\hline
		\end{tabular}
	\end{table}

\begin{figure}
  \includegraphics[width=\linewidth]{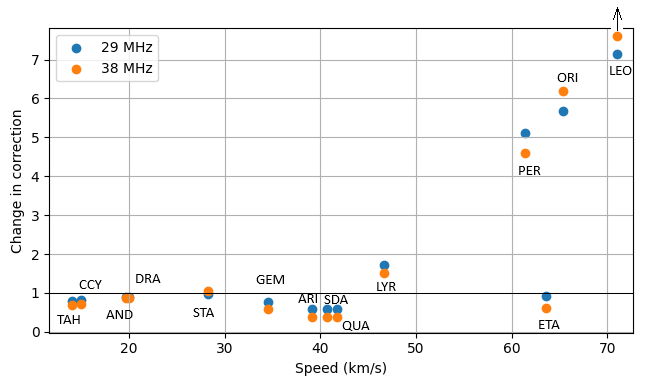}
  \caption[Shower correction change]{Plot of fractional change in shower correction factor; 1 means shower fluxes will not change, smaller numbers that fluxes will decrease }
  \label{fig:showers_changecor}
\end{figure}

In Fig.~\ref{fig:showers_changecor} we plot the change in correction factor for each shower, given by speed to make the trend clear. Table~\ref{tab:ShowCorFacs} shows the values. Note the new correction for the Leonids predicts an observed fraction so small, the change in flux we expect is a factor of 200 for 38 MHz. There is relatively little change for low speed showers, lower fluxes for midrange showers belonging partly or mostly to the lower population, and much higher fluxes for the fastest showers (except the ETAs). The fact that the Eta Aquariids are apparently much stronger or more refractory than other long-period comet showers is surprising, particularly since they come from comet 1P/Halley like the Orionids. This may be a function of the age of the shower \citep{Egal2020}, or a selection effect since the shower is only observable in darkness for a short time before dawn.

\section{Discussion}

\subsection{Comparison with current correction factors}

The new correction factors are similar to the old ones for the high population of meteors, which makes sense, since in most velocity bins this population is dominant for sporadics. It deviates at the highest speeds, which also makes sense, since the old initial radius correction was derived using amplitude ratios. Since the most severely affected echoes are not seen on both systems, there is a population which is not represented at all in the amplitude ratio data and therefore is not corrected for. The use of optical data allows us to correct this speed range properly for the first time. 

\subsection{Daytime showers}

We can only perform the $k_c$ analysis on nighttime showers, so for daytime showers we will have to choose a nighttime counterpart with similar properties. The $\beta$-Taurids and $\zeta$-Perseids are members of the Taurid complex, and will likely be similar to the STAs. The Daytime Sextantids are dynamically linked to the Geminids. The Arietids have a very low perihelion distance, and might therefore be similar to the SDAs. For now, this is the best we can do, though the outlier correction for the ETAs shows that this sort of analysis may not be correct.

\subsection{Overlap of radar and EMCCD sensitivity}\label{OverUnder}

One concern is whether these new correction factors apply to all the radar data. The EMCCD cameras are very sensitive, but their limiting magnitude is brighter than the limiting magnitude of the radar. One potential source of error is the proportion of overdense and transition echoes compared to underdense echoes. In the latter case, the radar beam penetrates the full width of the trail, so destructive interference is maximized when the trail is comparable in size to the radar wavelength. In overdense and transition echoes, there is a radiatively thick core to the trails which scatters radiation like a solid metal; in this case there will be much less destructive interference. More luminous meteors (roughly +5 and brighter) produce more ionization and denser electron clouds, so are more likely to be overdense, and therefore overrepresented in the brighter simultaneous optical/radar set. 

To investigate this issue, it is necessary to compare the simultaneously EMCCD/CMOR observed meteor population to that observed only by CMOR. To do this, we scanned the EMCCD data for dates and times when the meteor detection rate was high enough to clearly indicate relatively cloud-free conditions, selecting intervals where there was at least an hour of observing with at least 10 meteors/hour. While the CAMO system normally only operates when the sky is dark, the moon is below the horizon (or near new phase) and sky is clear \citep{Weryk2013}, there are occasionally conditions when some high cloud or thin fog may be present. 

This check should eliminate many marginal observing times with high cloud or fog. For those times, we searched the 29 and 38 MHz mpd files for echoes occurring in the fields of view of the EMCCDs, which could potentially have been detected by the cameras. We excluded 2017 and 2018 from this analysis, since the pointing changed very slightly during those years; this left 3172 hours of high-quality optical observations. We found 34,119 echoes on 29 MHz and 18,008 echoes on 38 MHz. 

We then looked at the raw amplitude-time profiles of these meteor echoes to determine whether they were underdense or not. In order to be considered underdense, an echo had to drop below the noise threshold in 108 pulses (approximately 0.2s) and not rise above two standard deviations over the noise for 2 seconds. These criteria are a bit crude; some short transition echoes may be misclassified as underdense, while some underdense echoes at low altitudes which have longer durations may be classified as overdense, but manual inspection suggests they are 90\% accurate at classifying echoes between underdense and overdense. 

\begin{figure}
  \includegraphics[width=\linewidth]{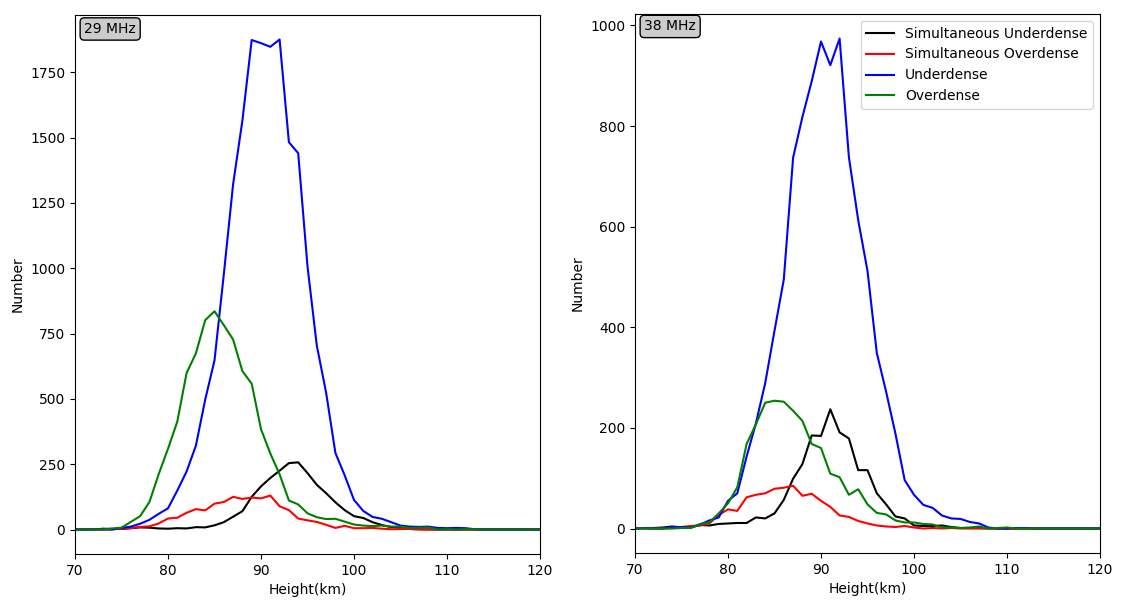}
  \caption[Distributions of under- and overdense meteors]{The absolute number of underdense and overdense echoes observed simultaneously with the EMCCD cameras (black and red) and by the radars in the EMCCD field of view (blue and green lines) during times when the cameras were running and the sky conditions were good.  }
  \label{fig:UnderOver}
\end{figure}

Figure~\ref{fig:UnderOver} shows the height distributions of the simultaneous optical and CMOR echoes in the EMCCD field of view and the total number of radar echoes during the time the EMCCDs were operating, sorted by underdense and overdense. In both cases, there is a larger fraction of underdense echoes in the full radar dataset (including faint meteors below the limiting sensitivity of the EMCCD cameras). The 29 MHz data show very significant differences between the simultaneous radar-EMCCD population and the full radar population.  There are many more echoes at lower heights not detected by the EMCCDs and both the underdense and overdense radar-only distributions peak at lower heights. The peaks in the 38 MHz data are similar in the simultaneous and full radar sets. This is not surprising since the 38 MHz system has a brighter limiting magnitude than the 29 MHz system, so a smaller fraction of echoes are missed, and those meteors are more similar to those at the lowest end of the EMCCD population.

It is more illuminating to look at the fraction of underdense and overdense echoes in each of the 6 speed bins for the two populations for 29 and 38 MHz. We cannot do the speed analysis for the larger set of CMOR echoes occurring in the EMCCD field of view but not detected by the EMCCDs, since CMOR's single station speed measurements based on the pre-t$_0$ phase approach \citep{Mazur2020} have significant uncertainty at our PRF in the highest speed bins. Thus we have to limit our speed analysis to the set of echoes also detected by EMCCDs, which provide the speeds. 

\begin{figure}
  \includegraphics[width=\linewidth]{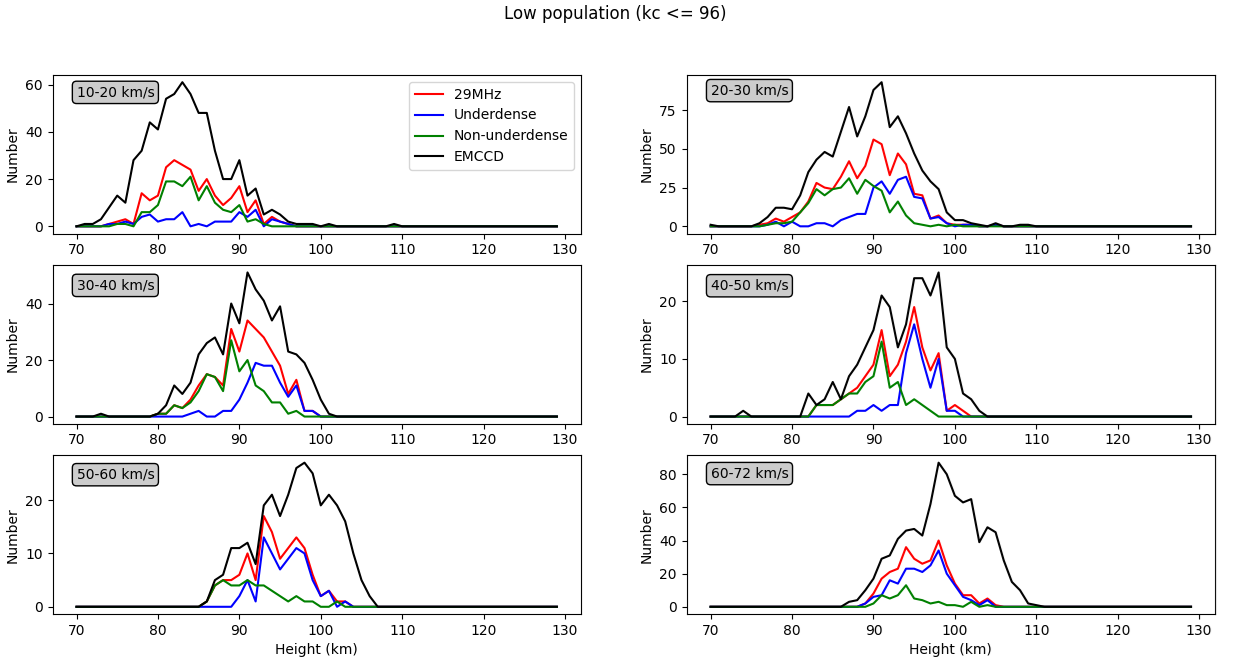}
  \caption[Distributions of under- and overdense meteors, 29 MHz, low population]{Distribution of underdense and overdense echoes observed simultaneously with 29 MHz and the EMCCD cameras, in each speed bin, for the low population.  }
  \label{fig:UnderOver29Low}
\end{figure}

\begin{figure}
  \includegraphics[width=\linewidth]{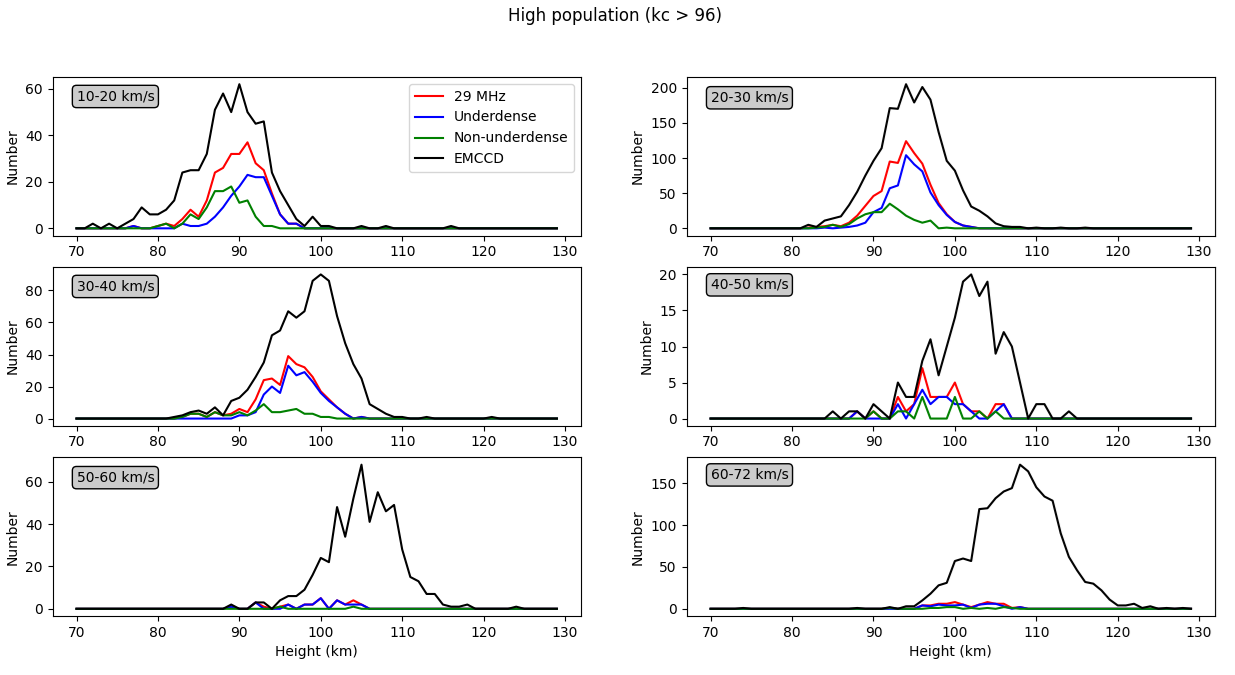}
  \caption[Distributions of under- and overdense meteors, 29 MHz, high population]{Distribution of underdense and overdense echoes observed simultaneously with 29 MHz and the EMCCD cameras, in each speed bin, for the high population.  }
  \label{fig:UnderOver29High}
\end{figure}

\begin{figure}
  \includegraphics[width=\linewidth]{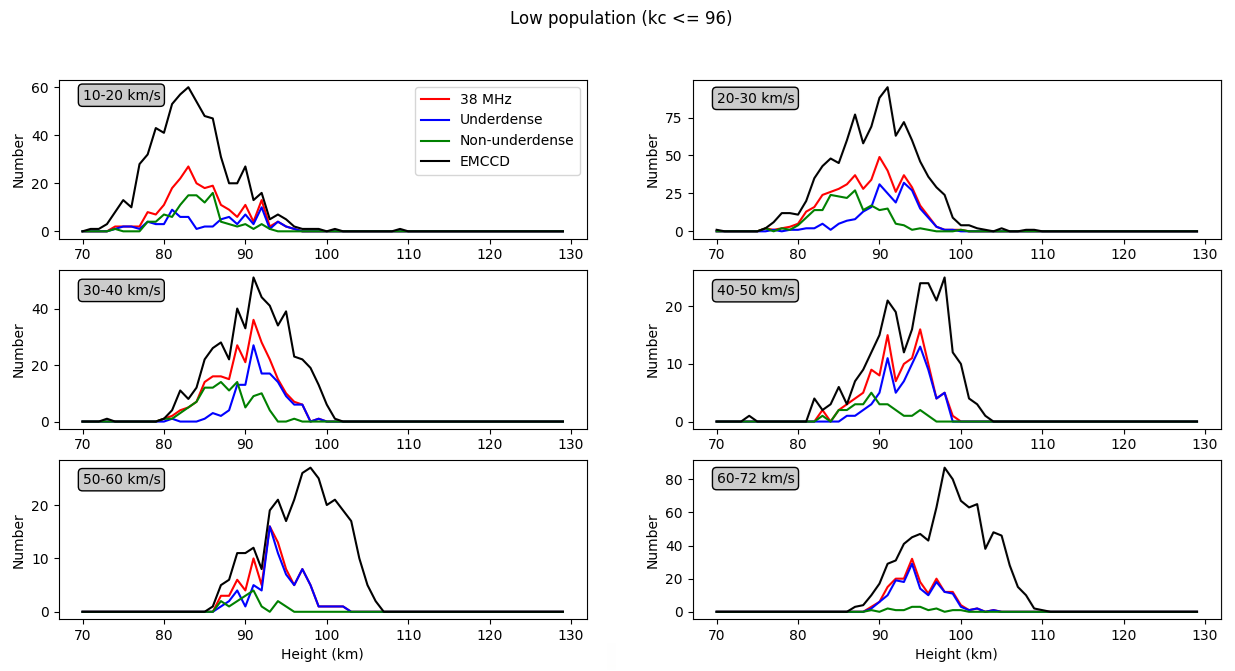}
  \caption[Distributions of under- and overdense meteors, 38 MHz, low population]{Distribution of underdense and overdense echoes observed simultaneously with 38 MHz and the EMCCD cameras, in each speed bin, for the low population.  }
  \label{fig:UnderOver38Low}
\end{figure}

\begin{figure}
  \includegraphics[width=\linewidth]{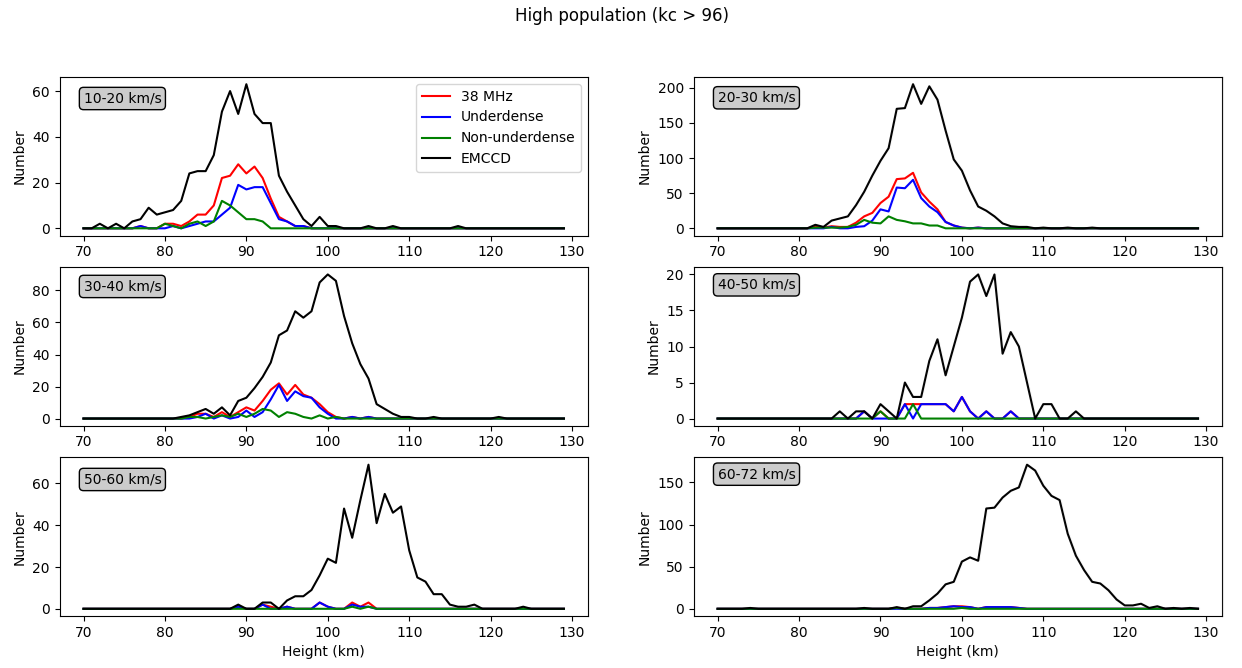}
  \caption[Distributions of under- and overdense meteors, 38 MHz, high population]{Distribution of underdense and overdense echoes observed simultaneously with 38 MHz and the EMCCD cameras, in each speed bin, for the high population.  }
  \label{fig:UnderOver38High}
\end{figure}

Figures~\ref{fig:UnderOver29Low} and \ref{fig:UnderOver29High} show the distribution for 29 MHz echoes, and Figures~\ref{fig:UnderOver38Low} and \ref{fig:UnderOver38High} show the 38 MHz distributions. In these plots the black line represents all the EMCCD meteors with two station solutions where the geometry is such that some portion of the lightcurve was specular relative to CMOR. The red line is the sum of all echoes detected at 29 (or 38) MHz which could be associated with an EMCCD event; the red line always lies below the black line as some EMCCD meteors are undetected by the radar  (because of low ionization, or destructive interference effects like initial radius, as described in the introduction). The radar echo numbers per bin are further broken down into morphologically overdense or underdense (blue and green coloured lines). The sum of these two coloured lines equals the red line in each plot - ie all meteor echoes can be classified as either underdense or overdense. 

In the highest speed bins, nearly all of the echoes are underdense, so the correction factors should hold for the fainter meteors missed by the EMCCD cameras. For the 38 MHz system, the high population has very few overdense echoes except at the lowest bin, and there the under and overdense echoes are both suppressed, presumably because the echoes are missed due to low electron line density, not from destructive interference in the scattering. 

There may be an issue with the 20-30 and 30-40 km/s bins for the low population, though 38 MHz still has many overdense echoes in the total population, so it is likely the correction factors are similar. The high population on 29 MHz is also mainly underdense except for the 20-30 km/s bin; again the lowest bin shows losses in both under and overdense meteors. The low population has many overdense echoes in the 20-30, 30-40, and 40-50 km/s bins, and the distribution of under and overdense echoes are quite different in the larger, fainter dataset. For this reason, a future task is to check the 29 MHz correction factors for mid-speed showers against the 38 MHz corrected fluxes to see if the 29 MHz correction in this paper is too small. 

The brightness bias of the simultaneous EMCCD/radar sample compared to all radar echoes is the most significant difference. There may also be a difference in the population of meteors which can be observed simultaneously with the current setup, given the geometry of the camera viewing volume and the echo lines for different radiants. Since the correction factor is not determined by the full population of meteors but rather for smaller bins in speed and (through the material properties) height, the corrections should be valid. We have a reasonable sample of meteors in each of the bins, and there is no reason to believe that (other than in size) the non-specular EMCCD meteors have different biases. 

\section{Conclusions}

We have produced new correction factors for the fraction of meteors not detected by specular meteor radars based on simultaneous observation (and non - observation) of a set of 10,503 EMCCD meteors with two station solutions having a portion of their optical trajectory at the specular point for the CMOR radar. During the time period that our cameras were recording we found some 34,119 and 18,008 meteor echoes in total occurred within the field of view of the EMCCD cameras at 29 and 38 MHz respectively. This demonstrates that a significant fraction of the specular radar echoes remain below the detection threshold of the EMCCD cameras. 

As expected, not all of the faint optical meteors were detected by the radar, despite almost all being well above the brightness detection threshold of the radar systems. This is mostly due to the attenuation effects which reduce specular scattering amplitudes, including the initial radius, finite velocity and PRF effects. At low speeds and heights, meteors are presumably missed because of low ionization. Another important correction to radar data is for Faraday rotation, which takes into account the attenuation of echoes from rotation of the polarization of the radar beam as it travels through the ionosphere. Since the ionosphere is only low enough to affect radar detections during the day, we cannot use our optical meteors to account for this effect, so its influence remains theoretical, based on classical theory \citep{Ceplecha1998}. Faraday rotation does not depend on the properties of the meteoroids producing radar echoes, so that theoretical correction is probably more accurate than others.

The new correction agrees reasonably well with the corrections currently in use for slow and medium speeds, for meteors which are weaker or more volatile and therefore part of the high population. The corrections will make the most difference to fast showers (for which the fluxes will increase significantly) and to moderate speed showers with strong or refractory meteoroids, which will decrease slightly. Most meteors belong to the high population, so it is not surprising that the \citet{Jones2005} correction agrees with our high population correction. At very high speeds, very few meteors are detected, so it's not surprising that there the correction was underestimated. 

The correction factors derived here are only applicable to the CMOR 29 and 38 MHz radars, or to other radars with similar wavelength and power. A radar with significantly lower transmitter power would see a higher fraction of overdense echoes, which suffer less attenuation from initial radius and finite velocity, while a higher power radar might need more correction to account for the increased fraction of underdense echoes at intermediate speeds. The PRF effect would also need to be investigated separately for radars with different pulse timing. At low speeds, where it appears a lack of ionization is responsible for the missed echoes, the correction factors probably depend less on the power and frequency, and are probably more generally applicable.

The effects of fragmentation on the radar amplitudes are included implicitly in these corrections. In the future, high resolution CAMO tracking images of meteors at the specular point could be used to explicitly investigate the role of fragmentation on individual multi-frequency echoes, using a scattering model.

In a future paper, we will compare the fluxes of individual showers on the two frequencies to verify that the corrections produce the same flux for both systems. This requires additional work to verify limiting magnitudes and mass indices for the showers. We will also investigate any effects of zenith angle of the radiant or mass index on the correction factors.  

\section{Acknowledgements}

This work was supported in part by the NASA Meteoroid Environment Office under cooperative agreement 80NSSC24M0060 and by the Natural Sciences and Engineering Research council of Canada.

% Import the bibliography file main.bib
\bibliography{main}

\newpage

\end{document}